\documentclass[12pt,draftclsnofoot,onecolumn]{IEEEtran} \usepackage{amssymb}

\usepackage{blindtext}
\usepackage{amssymb}
\usepackage{amsmath}
\usepackage{graphicx}
\usepackage{cite}
\usepackage{citesort}
\usepackage{multicol}
\usepackage{amsfonts}
\usepackage{color}
\usepackage{subfigure}
\usepackage{bm}
\usepackage{epstopdf}
\usepackage{amssymb}
\usepackage{comment}
\usepackage{amsmath}
\allowdisplaybreaks[4]
\usepackage{booktabs}
\usepackage{multirow}
\usepackage{bmpsize}
\usepackage[ruled,linesnumbered]{algorithm2e}

\usepackage{lipsum}
\usepackage{stfloats}

\newtheorem{theorem}{\textbf{Theorem}}

\newtheorem{lemma}{\textbf{Lemma}}

\newtheorem{corollary}{\textbf{Corollary}}

\makeatletter
\def\ScaleIfNeeded{%
\ifdim\Gin@nat@width>\linewidth \linewidth \else \Gin@nat@width
\fi } \makeatother

\begin{document}
%

\title{{Edge Intelligence in Satellite-Terrestrial Networks with Hybrid Quantum Computing}}

\author{Siyue Huang, Lifeng Wang, Xin Wang, Bo Tan, Wei Ni and Kai-Kit Wong
\thanks{S. Huang, L. Wang and X. Wang are with the School of Information Science and Engineering, Fudan University, Shanghai 200433, China (e-mail: $\rm \{lifengwang,xwang11\}@fudan.edu.cn$).}
\thanks{B. Tan is with the Faculty of Information Technology and Communication Sciences, Tampere University, Finland (E-mail: $\rm bo.tan@tuni.fi$).}
\thanks{ W. Ni is with the Commonwealth Scientific and Industrial Research Organization (CSIRO), Sydney, NSW 2122, Australia (e-mail: wei.ni@data61.csiro.au).}
\thanks{K.-K. Wong is with the Department of Electronic and Electrical Engineering, University College London, London WC1E 7JE, U.K.; K.-K. Wong is also affiliated with Yonsei Frontier Lab, Yonsei University, Korea  (e-mail:
kai-kit.wong@ucl.ac.uk).}
}


\maketitle

\begin{abstract}
This paper exploits the potential of edge intelligence empowered satellite-terrestrial networks, where users' computation tasks are offloaded to the satellites or terrestrial base stations. The computation task offloading in such networks involves the edge cloud selection and bandwidth allocations for the access and backhaul links,  which aims to minimize the energy consumption under the delay and satellites' energy constraints. To address it, an alternating direction method of multipliers (ADMM)-inspired algorithm is proposed to decompose the  joint optimization problem into small-scale subproblems. Moreover, we develop a hybrid quantum double deep Q-learning (DDQN)  approach to optimize the edge cloud selection. This novel deep reinforcement learning architecture enables that classical and quantum neural networks process information in parallel. Simulation results confirm the efficiency of the proposed algorithm, and indicate that duality gap is tiny and a larger reward can be generated from a few data points compared to the classical DDQN.
\end{abstract}


\section{Introduction}
Low-carbon economy incentivizes future 6G networks to be more environment-friendly. However, edge intelligence with machine learning algorithms may lead to more energy consumptions in  edge computing-enabled terrestrial networks~\cite{Xiaoyan2020}. Space computing in satellite networks is a promising approach to reducing energy costs since the satellites harvest solar energy~\cite{Alan-D}.  With the development of dense  satellite constellations~\cite{Furong2023}, space computing resources can be abundant, which need to be efficiently utilized.

 The implementation of machine learning for managing the radio resources and edge computing resources of satellite networks has been studied in the literature~\cite{Hassan_JSAC2024,Hangyu2024}. The data packet routing  problem in satellite constellations is investigated in \cite{Hassan_JSAC2024}, which is solved by employing the deep reinforcement learning (DRL). The work~\cite{Hangyu2024} studies the  satellite cooperative computing and proposes a multi-agent collaborative
 task offloading scheme. Meanwhile, recent progress in quantum machine learning opens up a new research avenue~\cite{PennyLane}.  The variational quantum deep Q-learning with less memory consumption and fewer neural network model parameters is designed in \cite{Yen-Chi2020}, which is demonstrated in  cognitive radio environment. In \cite{Narottama2022},  a quantum neural network is leveraged to make user grouping, which reduces the complexity compared to the classical neural networks. The results in \cite{Rainjonneau2023} show that a quantum machine learning algorithm can deal with satellite mission planning problem more efficiently than the classical ones. In the edge computing-based IoT systems, \cite{Ansere2024} proposes a quantum deep  Q-learning  scheme to improve the content delivery efficiency.

Computation task offloading plays an essential role in the 6G satellite-terrestrial networks, where both the terrestrial base stations (BSs) and satellites can be edge cloud servers. Although the BSs have  stable energy supply, they may consume non-renewable energy resources and give rise to more carbon emissions when a large variety of applications such as extended reality (XR)~\cite{huming} generate massive computation-intensive tasks. On the other hand, the satellites depend on the limited  energy harvested from solar panels~\cite{yang2016JSAC}, and recent work~\cite{Feng-Wei} underscores that energy management for edge intelligence at the satellites is critical. Therefore, both the latency requirement and energy constraint need to be met when tackling the task offloading issue, to make edge intelligence at the satellites sustainable.

Motivated by the aforementioned studies, this paper focuses on  energy-efficient task offloading in  satellite-terrestrial networks. Considering that the latency for edge intelligence and satellites' energy are stringently restricted, we seek to minimize the total energy consumption through optimizing the edge cloud selection and bandwidth allocation.




\section{System Descriptions}\label{System_description}
We consider a general satellite-terrestrial network consisting of $J-1$ cooperative satellites located in the same/different orbits and one terrestrial BS indexed by $J$. The satellites in the same orbit may have  identical computing capability. Inter-satellite link (ISL) is leveraged to support  cooperative edge computing between the satellites. The computation tasks offloaded by multiple user equipment (UEs) are executed at the edge cloud servers including the satellites or terrestrial BS, which can save the UEs' energy and mobile computing resources. Suppose that  $N$ UEs in a terrestrial millimeter wave (mmWave) cell send their computation tasks to the BS, the access transmission latency $T_{\rm access}^n$ for the $n$-th UE is given by
\begin{align}\label{eq1}
& T_{\rm access}^n  = \frac{{I_n }}{{R_{\rm access}^n }} \nonumber\\
&=\frac{{I_n }}{\sum\limits_{j = 1}^J {x_{n,j} B_{\rm access}^{n,j} \log _2 \left( {1 + \frac{{G_{\rm UE}^{\rm Tx} G_{\rm BS}^{\rm Rx} p_{\rm UE}^n }}{{B_{\rm access}^{n,j} \delta_{\rm a}^2 }}\left| {h_{\rm access}^n } \right|^2 } \right)} },
\end{align}
where $I_n$ is the number of bits for the $n$-th UE's computation task; $R_{\rm access}^n$ is the access transmission rate; $x_{n,j}$ is the binary association indictor, namely, $x_{n,j}=1$ denotes that the UE's computation task is  offloaded to the $j$-th edge cloud server; $B_{\rm access}^{n,j}$ is the allocated access frequency bandwidth at the UE $n$;  $G_{\rm UE}^{\rm Tx}$ and $G_{\rm BS}^{\rm Rx}$ are the UE's effective transmit antenna gain and BS's effective receive antenna gain, respectively; $p_{\rm UE}^n$ is the $n$-th UE's transmit power;  $\delta_{\rm a}^2$ is the noise's  power spectral density (PSD);  $\left| {h_{\rm access}^n } \right|^2$ is the large-scale fading channel power gain.

After receiving these computation tasks, the BS may deliver some of them to its closest satellite, and its closest satellite may  proceed to forward the computation tasks to its cooperative satellites via ISL. When offloading the $n$-th UE's computation task to the $j$-th satellite, the total transmission latency  is given by
\begin{align}\label{eq2}
& T_{\rm Sat}^{n,j}  =
T_{\rm access}^n  + \underbrace {\frac{{I_n }}{{R_{\rm Sat}^n }} + \frac{{d_{\rm Sat} }}{c}+
\frac{{I_n }}{{R_{\rm ISL}^n }}{\rm H}_{\rm ISL}^j  + \tau _{\rm ISL}^j }_{\rm backhaul~latency},
\end{align}
where $T_{\rm access}^n$ is given in \eqref{eq1}; $R_{\rm Sat}^n $ and $d_{\rm Sat}$ are the backhaul transmission rate and communication distance from the BS to its closest satellite for delivering UE $n$'s task, respectively; $c$ is the electromagnetic wave's speed; $R_{\rm ISL}^n$ is the ISL's transmission rate;  ${\rm H}_{\rm ISL}^j$ and $\tau _{\rm ISL}^j$ are the number of hops and propagation delay from the BS's closest satellite to the targeted satellite $j$, respectively; $R_{\rm Sat}^n $  is
\begin{align}\label{eq3}
\hspace{-0.2cm} R_{\rm Sat}^n  = \sum\limits_{j = 1}^{J - 1} {x_{n,j} B_{\rm S}^{n,j} } \log _2 \left( {1 + \frac{{G_{\rm BS}^{\rm Tx} G_{\rm Sat}^{\rm Rx} p_{\rm BS}^n }}{{B_{\rm S}^{n,j} \delta_{\rm S}^2 }}\left| {h_{\rm Sat} } \right|^2 } \right),
\end{align}
where $B_{\rm S}^{n,j} $ denotes the allocated bandwidth at the BS for delivering the offloaded computation task to the $j$-th satellite; $G_{\rm BS}^{\rm Tx}$ and $G_{\rm Sat}^{\rm Rx}$ are the BS's effective transmit antenna gain and the satellite's receive antenna gain, respectively; $p_{\rm BS}^n $ is the BS's transmit power for the $n$-th UE's task data; $\delta_{\rm S}^2$ is the noise's  PSD; $\left| {h_{\rm Sat} } \right|^2$ is the large-scale channel power gain between the BS and the satellite.

 Suppose that equal computing resource allocation is adopted at each edge cloud server, the total energy consumption for the $n$-th UE's task offloading and computation at the BS is
 \begin{align}\label{eq4}
E_{\rm terr}^n\left(\mathbf{x},\mathbf{B}_{\rm access}\right) = \frac{{I_n }}{{R_{\rm access}^n }}p_{\rm UE}^n  + \eta _{\rm terr}^n  I_n \kappa_n \left(\frac{f_{\rm terr} }{\sum\limits_{n = 1}^N {x_{n,J}}} \right)^2,
\end{align}
where $\mathbf{x}=[x_{n,j}]$; $\mathbf{B}_{\rm access}=[B_{\rm access}^{n,j}]$; $\eta_{\rm terr}^n$ is the effective switched capacitance of the BS; $\kappa_n$ (CPU cycles/bit) is the amount of required computing resources for computing 1-bit
 of the offloaded data~\cite{Xiaoyan2020}; $f_{\rm terr}$ is the total CPU clock frequency of the BS.

 Since the satellites sustain themselves and generate the electricity from the solar panels, the total energy consumption for the $n$-th UE's task offloading and computation at the targeted satellite is given by
  \begin{align}\label{eq5}
E_{\rm Sat}^n  = \frac{{I_n }}{{R_{\rm access}^n }}p_{\rm UE}^n  + \frac{{I_n }}{{R_{\rm Sat}^n }} p_{\rm BS}^n.
\end{align}

 Our objective is to minimize the total energy consumption of the satellite-terrestrial network, which is given by
\begin{align}\label{eq6}
& \mathop {\min }\limits_{\mathbf{x},\mathbf{B}}~ \underbrace{\sum\limits_{n = 1}^N \sum\limits_{j = 1}^{J - 1} x_{n,j} E_{\rm Sat}^n} _{E^{\rm total}_{\rm Sat}\left(\mathbf{x},\mathbf{B}\right)}  +\sum\limits_{n = 1}^N x_{n,J} E_{\rm terr}^n\left(\mathbf{x},\mathbf{B}\right)   \\
&\mathrm{s.t.}~ \mathrm{C1}:\;x_{n,j} \left( {1 - x_{n,j} } \right) = 0,\quad \forall n,j, \nonumber\\
 &\mathrm{C2}:\;\sum\limits_{j = 1}^J {x_{n,j}  = 1},\quad \forall n, \nonumber \\
 &\mathrm{C3}:\;\sum\limits_{j = 1}^{J - 1} x_{n,j} \underbrace {\left(T_{\rm Sat}^{n,j}+\frac{I_n \kappa_n}{f_{\rm Sat}^{j}} \sum\limits_{n = 1}^N {x_{n,j}}\right)}_{T_{\rm Sat,total}^{n,j}} \nonumber \\
 &\quad\quad\quad\quad \quad+ x_{n,J} \underbrace {\left(T_{\rm access}^n+\frac{I_n \kappa_n}{f_{\rm terr}}\sum\limits_{n = 1}^N {x_{n,J}}\right)}_{T_{\rm Terr,total}^{n}}  \le T_{\rm th},~ \forall n, \nonumber \\
&\mathrm{C4}:\;\sum\limits_{n = 1}^N {\sum\limits_{j = 1}^J {x_{n,j} B_{\rm access}^{n,j} }  } \le  B_{\rm access}^{total}, \nonumber \\
&\mathrm{C5}:\;\sum\limits_{n = 1}^N {\sum\limits_{j = 1}^{J - 1} {x_{n,j} B_{\rm S}^{n,j} } }  \le B_{\rm S}^{total},\quad \nonumber \\
&\mathrm{C6}:\;\sum\limits_{n = 1}^N x_{n,j} \eta_{\rm Sat}^n I_n \kappa _n \left( \frac{f_{\rm Sat}^{j}}{\sum\limits_{n = 1}^N {x_{n,j}}} \right)^2   \le E_{\rm th}^j,\forall j=1,\cdots, J-1, \nonumber\\
&\mathrm{C7}: \; B_{\rm access}^{n,j}\geq 0,~B_{\rm S}^{n,j}  \geq 0, \quad \forall n,j,\nonumber
\end{align}
 where $\mathbf{B}=[B_{\rm access}^{n,j},B_{\rm S}^{n,j}]$; $f_{\rm Sat}^{j}$ is the total CPU clock frequency of the satellite $j$; $T_{\rm Sat,total}^{n,j}$ and $T_{\rm Terr,total}^{n}$ represent the sum of transmission latency and the edge computing latency at the satellite and BS, respectively;  $\eta_{\rm Sat}^n$ is the effective switched capacitance of the satellite.

 In problem \eqref{eq6}, constraints $\mathrm{C1}$--$\mathrm{C2}$ make sure that each UE is solely served by one edge cloud server;  constraint $\mathrm{C3}$ is the maximum allowable latency for edge computing;  $\mathrm{C4}$--$\mathrm{C5}$ are the constrained frequency resources for the access and backhaul; constraint $\mathrm{C6}$ illustrates that the satellite's energy is limited, which depends on many factors including the orbital plane and eclipse~\cite{yang2016JSAC}, particularly the limited energy harvested by a low Earth orbit (LEO) for one orbit period~\cite[Lemma 1]{yuan2023}. By introducing the auxiliary variable vector $\bm{\xi}=[\xi_{j^{'}}]$ ($j^{'}=1,\cdots,N\left(2J-1\right)$), problem \eqref{eq6} is  equivalently transformed as
 \begin{align}\label{eq7}
& \hspace{-0.2cm} \mathop {\min }\limits_{\{\mathbf{x},\mathbf{B},\bm{\xi}\}\in \mathcal{X}}~E_{\rm total}\left(\mathbf{x},\mathbf{B},\bm{\xi}\right)=E^{\rm total}_{\rm Sat}\left(\mathbf{x},\mathbf{B}\right)+ \sum\limits_{n = 1}^N x_{n,J} E_{\rm terr}^n\left(\mathbf{x},\bm{\xi}\right) \nonumber \\
&\mathrm{s.t.}~\mathbf{B}=\bm{\xi},
\end{align}
where the constraint set $\mathcal{X}$ is defined as
\begin{align}\label{eq8}
\mathcal{X}=\left\{\left(x_{n,j},B_{\rm access}^{n,j},B_{\rm S}^{n,j},\xi_{j^{'}}\right)|\mathrm{C1}-\mathrm{C7}, \xi_{j^{'}} \geq 0\right\}.
\end{align}
The introduction of auxiliary variables enables that problem \eqref{eq6} can be split into multi-block separable problems and is leveraged to construct strong convexity for splitting algorithm design at next section.

The sharing problem \eqref{eq7} is  non-convex, and the scheduling parameter and frequency allocation are coupled. To efficiently address it, we propose a novel  algorithm with hybrid quantum computing in the following section.
\section{Splitting Algorithm Design}\label{HQC}
The sharing problem \eqref{eq7} needs to be properly solved in the considered delay-limited networks.  To reduce the computational complexity, the alternating direction method of multipliers (ADMM) inspired algorithm is developed. ADMM has been adopted to efficiently solve non-convex problems~\cite{zhiquanL2016} and mixed-integer programming~\cite{laudecvpr2018}.  Hence, the augmented Lagrangian with respect to (w.r.t.) problem \eqref{eq7} is given by
\begin{align}\label{eq9}
&\mathcal{L}\left(\mathbf{x},\mathbf{B},\bm{\xi},\bm{\varpi}\right)= E_{\rm total}\left(\mathbf{x},\mathbf{B},\bm{\xi}\right)+\frac{\rho}{2}\left\|\mathbf{B}-\bm{\xi}-\bm{\varpi}\right\|_2^2, 
\end{align}
where $\rho>0$ is the penalty parameter, and $\bm{\varpi}=[\varpi_n]$ is the scaled dual variable vector. Thus problem \eqref{eq7} is decomposed into small-scale subproblems at each iteration, namely
 \begin{align}
&\mathbf{x}^{(\ell+1)}=\mathop {\rm argmin}\limits_{ \mathbf{x} \in \mathcal{X} } \mathcal{L}\left(\mathbf{x},\mathbf{B}^{(\ell)},\bm{\xi}^{(\ell)},\bm{\varpi}^{(\ell)}\right), \label{eq10} \\
&\mathbf{B}^{(\ell+1)}=\mathop {\rm argmin}\limits_{ \mathbf{B} \in \mathcal{X} } \mathcal{L}\left(\mathbf{x}^{(\ell+1)},\mathbf{B},\bm{\xi}^{(\ell)},\bm{\varpi}^{(\ell)}\right), \label{eq11} \\
&\bm{\xi}^{(\ell+1)}=\mathop {\rm argmin}\limits_{ \bm{\xi} \in \mathcal{X} } \mathcal{L}\left(\mathbf{x}^{(\ell+1)},\mathbf{B}^{(\ell+1)},\bm{\xi},\bm{\varpi}^{(\ell)}\right), \label{eq13} \\
&\bm{\varpi}^{(\ell+1)}=\bm{\varpi}^{(\ell)}-\Bigg(\mathbf{B}^{(\ell+1)}-\bm{\xi}^{(\ell+1)}\Bigg),  \label{eq14}
\end{align}
where $\ell$ is the iteration index. 

\begin{figure}[t!]
\centering
\includegraphics[width=3.5 in]{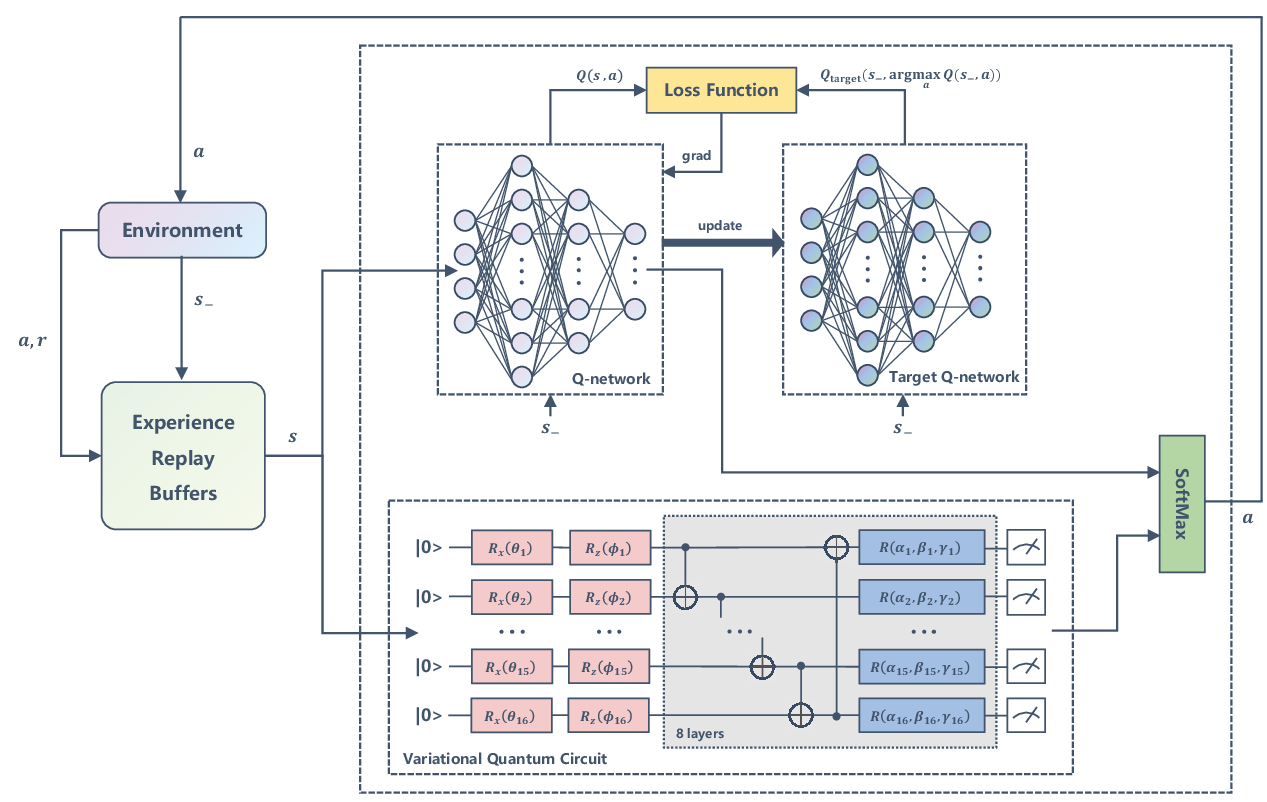}
\caption{Hybrid quantum DDQN architecture with classical and quantum deep Q-networks, where $R_x\left(\theta\right)$ and $R_x\left(\phi\right)$ with the rotations along x-axis
and z-axis by the angles $\theta$ and $\phi$ are the single-qubit gates, and $R\left(\alpha,\beta,\gamma\phi\right)$ is the general single qubit unitary gate in the quantum model~\cite{Yen-Chi2020}.
}
\label{DRL_Quantum}
\end{figure}

The solution of the non-convex constrained subproblem \eqref{eq10} can be well approximated with safety guarantees  using the  primal-dual algorithm method with DRL~\cite{Santiago2023}, in particular, \cite{Santiago2023} shows that the duality gap can be minimal when the neural network employed by DRL~\cite{huming2024} has the sufficiently rich parametrization. Therefore, the dual function of  \eqref{eq10} is
{\begin{align}\label{eq15}
&\hspace{-0.2cm} d\left(\bm{\lambda},\bm{\bar{\lambda}},\varphi,\psi,\bm{\mu}\right)=\mathop {\min }\limits_{\mathbf{x}\in \{0,1\}} E_{\rm total}\left(\mathbf{x},\mathbf{B},\bm{\xi}\right)+\sum\limits_{n = 1}^{N} \lambda_n \left(\sum\limits_{j = 1}^J x_{n,j}-1 \right)\nonumber\\
&\hspace{-0.1cm} + \sum\limits_{n = 1}^{N} \bar{\lambda}_n \left(\sum\limits_{j = 1}^{J - 1} x_{n,j}^{(\ell+1)}T_{\rm Sat,total}^{n,j}+ x_{n,J}^{(\ell+1)}T_{\rm Terr,total}^{n} -T_{\rm th}\right) \nonumber\\
&+ \varphi\left(\sum\limits_{n = 1}^N {\sum\limits_{j = 1}^J {x_{n,j} B_{\rm access}^{n,j} }  }- B_{\rm access}^{total}\right)
+\psi\Bigg(\sum\limits_{n = 1}^N {\sum\limits_{j = 1}^{J - 1} {x_{n,j} B_{\rm S}^{n,j} } }   \nonumber\\
&\hspace{-0.1cm}- B_{\rm S}^{total}\Bigg)+ \sum\limits_{j = 1}^{J-1} \mu_j \left(\sum\limits_{n = 1}^N x_{n,j} \eta_{\rm Sat}^n I_n\kappa _n \left(  \frac{f_{\rm Sat}^{j}} {\sum\limits_{n = 1}^N x_{n,j}}\right)^2   - E_{\rm th}^j\right),
\end{align}}
where $\bm{\lambda}=[\lambda_n]$ is the dual variable vector; $\bm{\bar{\lambda}}=[\bar{\lambda}_n]$, $\varphi$, $\psi$ and $\bm{\mu}=[\mu_j]$  are the positive dual variables.
Although the conventional double deep Q-learning (DDQN)~\cite{Hado2016} can solve the discrete problem,  efficiently computing the dual function \eqref{eq15} may require rich enough parameterizations since it involves the constrained non-convex problem~\cite{Santiago2023}. To this end, we propose a hybrid quantum DDQN solution as illustrated in Fig. \ref{DRL_Quantum}. Compared with the conventional counterpart, the benefits of this new solution are twofold: i) By integrating the classical and quantum neural networks in a parallel manner, classical neural network's parameter dimension  and computation complexity are decreased; ii) The quantum model consisting of variational quantum circuits usually helps generalize larger reward from a few data points~\cite{Rainjonneau2023}, which is also seen in the results of Section~\ref{sec:simulation} at next page. When applying the proposed hybrid quantum DDQN to compute the dual function \eqref{eq15} (its negative value is referred to as reward) for fixed dual variables, the agent (namely BS) interacts with the environment. Let $\emph{\textbf{s}}$ and $\emph{\textbf{s}}_{-}$  denote the agent's current and next states, respectively, including all the link conditions, bandwidth allocations, and transmit powers; the  agent's action $\emph{\textbf{a}}$ represents the association decisions. Differing from the classical deep Q-network, in this work, hybrid quantum deep Q-network evaluates the action-values (i.e., the Q values) $\mathbf{Q}_{\rm hybrid}\left(\emph{\textbf{s}},\emph{\textbf{a}}\right)$ by combining the outputs of both classical and quantum deep Q-networks, i.e.,
\begin{align}\label{eq16}
\hspace{-0.3cm}\mathbf{Q}_{\rm hybrid}\left(\emph{\textbf{s}},\emph{\textbf{a}}\right)={\rm diag}(\emph{\textbf{w}}_{\rm c}) \mathbf{Q}_{\rm c}\left(\emph{\textbf{s}},\emph{\textbf{a}}\right)+{\rm diag}(\emph{\textbf{w}}_{\rm q})\mathbf{Q}_{\rm q}\left(\emph{\textbf{s}},\emph{\textbf{a}}\right),
\end{align}
where $\mathbf{Q}_{\rm c}$ and $\mathbf{Q}_{\rm q}$ are the Q values from the classical and quantum deep Q-networks, respectively, ${\rm diag}\left(\textbf{w}\right)$ denotes the diagonal matrix with the diagonal elements contained by the vector $\textbf{w}$;  $\emph{\textbf{w}}_{\rm c}$ and $\emph{\textbf{w}}_{\rm q}$ are the trainable parameters of the hybrid quantum deep Q-network. After computing the dual function \eqref{eq15}, its corresponding dual problem w.r.t. dual variables is convex and can be  solved via gradient algorithm. Thus the subproblem \eqref{eq10} is efficiently addressed.

Given $\mathbf{x}$, the subproblem \eqref{eq11} is  a convex problem w.r.t. $\mathbf{B}$, and thus Karush-Kuhn-Tucker (KKT) condition can be adopted to solve it since the Slater's condition holds.  The subproblem \eqref{eq13} is the  convex problem, moreover, it can be split into $N\left(2J-1\right)$ subproblems and computed in a parallel manner. Thus the solutions of the subproblems (10)--(12) at each iteration are obtained. To guarantee the convergence, the descent condition that the augmented Lagrangian value decreases monotonically with the iterates  is met, namely
\begin{align}\label{eq17}
\hspace{-0.3cm}\mathcal{L}\left(\mathbf{x}^{(\ell+1)},\mathbf{B}^{(\ell+1)},\bm{\xi}^{(\ell+1)},\bm{\varpi}^{(\ell)}\right) - \mathcal{L}\left(\mathbf{x}^{(\ell)},\mathbf{B}^{(\ell)},\bm{\xi}^{(\ell)},\bm{\varpi}^{(\ell)}\right) \leq \varepsilon,
\end{align}
where $\varepsilon$ is the non-positive value.
\vspace{-0.3cm}

\section{Simulation Results}\label{sec:simulation}
In our simulations, there are four UEs with the same size of computation tasks served by the BS and their communication distances are uniformly distributed with the interval $[100,400]$ m; there are three LEO satellites located in the same orbit with the altitude 600 km and the middle one is connected to the BS at nadir, the ISL's propagation delay is $\tau _{\rm ISL}=1.46$ ms, and the other basic parameters are shown in Table I.
\vspace{-0.3cm}
{\begin{flushleft}
\begin{table}[h]\label{table1} \footnotesize
\centering
\caption{Simulation parameters}
\setlength\tabcolsep{1.2 pt}
\hspace{-0.7cm} \begin{tabular}{|l|l|}
  \hline
    Effective transmit antenna gain per UE  & $G_{\rm UE}^{\rm Tx}=4$dBi  \\ \hline
  BS's effective receive and transmit antenna gain   & $G_{\rm BS}^{\rm Rx}=15$dBi;~$G_{\rm BS}^{\rm Tx}=38$dBi  \\ \hline
  Satellite's  effective receive antenna gain & $G_{\rm Sat}^{\rm Rx}=38$dBi \\ \hline
  Transmit power per UE & $p_{\rm UE}^n=23$dBm, $\forall n$ \\ \hline
  BS's transmit power per UE's data stream & $p_{\rm BS}^n=40.97$dBm, $\forall n$ \\ \hline
  mmWave carrier frequency (CF) for access links &  $f_{\rm CF}^{\rm a}= 28$GHz  \\ \hline
 mmWave CF for satellite-terrestrial backhaul link &  $f_{\rm CF}^{\rm b}=30$GHz  \\ \hline
   Large-scale channel fading power gain &  $\left|\hbar_n\right|^2=\left( {\frac{3\times 10^8}{{4\pi f_{\rm CF} }}} \right)^2 \times d^{ - 2} $   \\ \hline
Effective switched capacitance of the BS & $\eta_{\rm terr}^n=10^{-28}$, $\forall n$  \\ \hline
Required CPU cycles per bit & $\kappa_n=300$, $\forall n$ \\ \hline
Total CPU clock frequency of the BS & $f_{\rm terr}=3\times 10^{9}$ \\ \hline
Total CPU clock frequency of the satellite & $f_{\rm Sat}^{j}=3\times 10^{9}$, $\forall j$ \\ \hline
Maximum allowable latency & $T_{\rm th}=0.105$s \\ \hline
The maximum available energy at the satellite & $E_{\rm th}^j=0.5$, $\forall j$, \\ \hline
ISL's transmission rate & $R_{\rm ISL}^n=10$Gbps \\ \hline
\end{tabular}
\end{table} \vspace{-0.3cm}
\end{flushleft}
}
\vspace{-0.1cm}
\subsection{Convergence}\label{converg}
We first compare the proposed hybrid quantum DDQN architecture with its classical counterpart. Specifically, the classical Q-network is the fully connected multi-layered perceptron including the input layer, output layer and two hidden layers, where the first and second hidden layers respectively have 256 and 128 neurons as depicted in Fig. 1. The proposed hybrid quantum Q-network adopts the small classical and quantum Q-network, where the small classical Q-network includes two hidden layers of sizes 64 and 32 neurons, and the quantum Q-network is the 16-qubit parametrized quantum circuit~\cite{Rainjonneau2023}. In the simulations, we adopt the PennyLane library in the Python software environment for hybrid quantum computing~\cite{PennyLane}.
\begin{figure}
     \centering
    \subfigure[{{$B_{\rm access}^{total}=50$}}MHz,$B_{\rm S}^{total}=100$MHz.]{
         \centering
         \includegraphics[width=1.6 in,height=1.5 in]{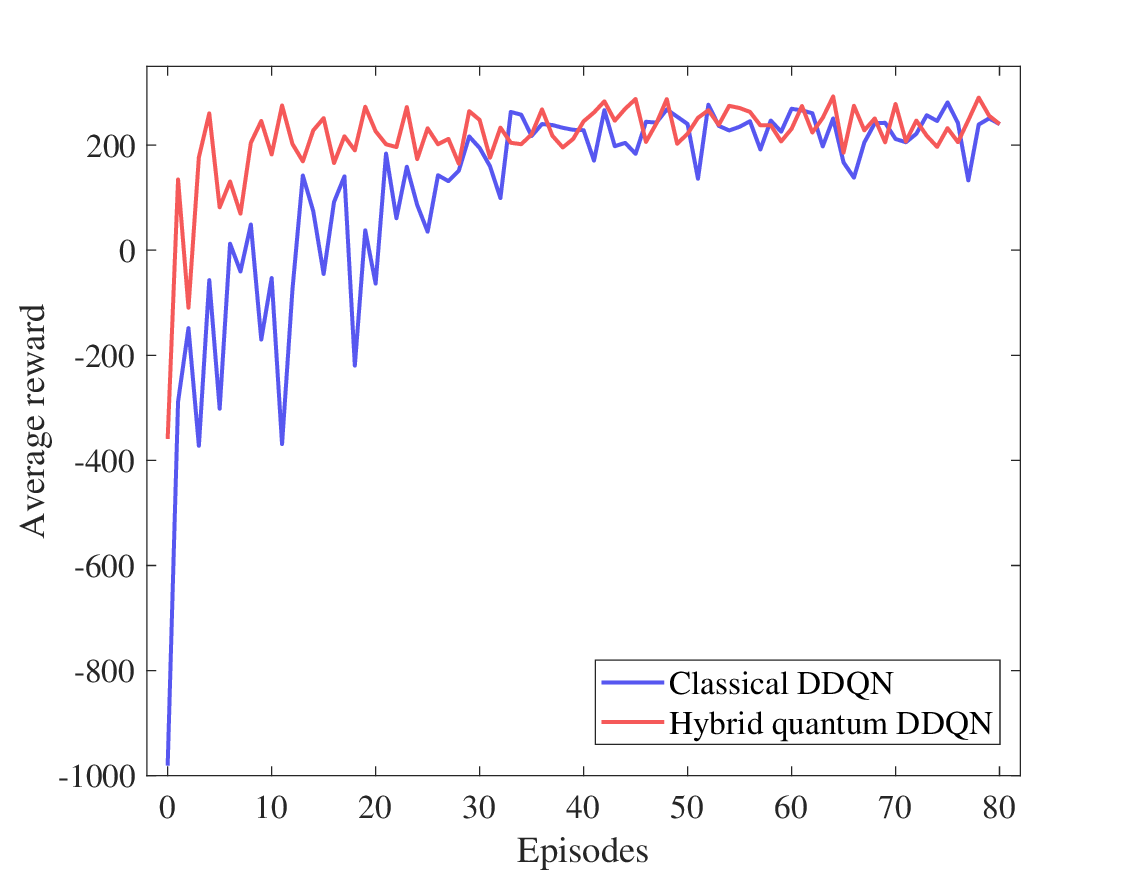}
      \label{figa1_APc}
     }
     \subfigure[{{$B_{\rm access}^{total}=60$}}MHz,$B_{\rm S}^{total}=100$MHz.]{
         \centering
         \includegraphics[width=1.6 in,height=1.5 in]{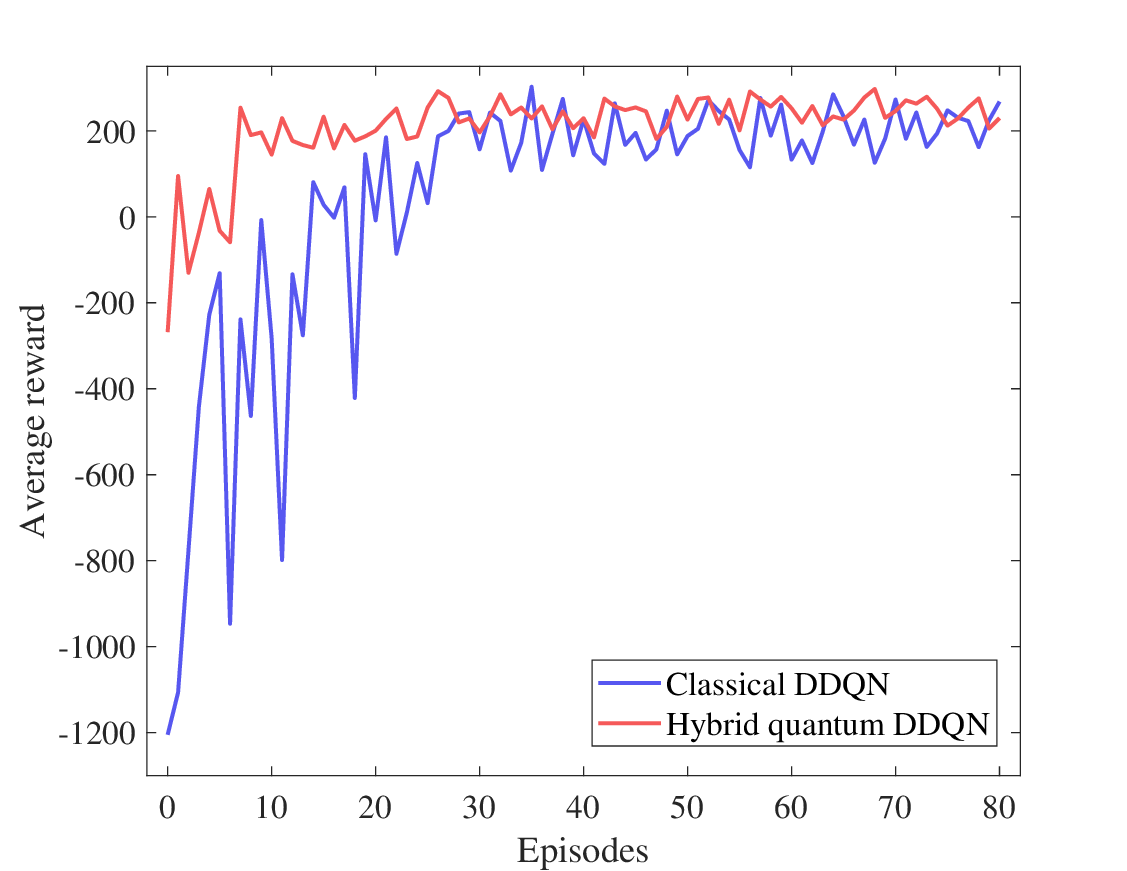}
      \label{figb1_APc}
    }
         \subfigure[{{$B_{\rm access}^{total}=50$}}MHz,$B_{\rm S}^{total}=110$MHz.]{
         \centering
         \includegraphics[width=1.6 in,height=1.5 in]{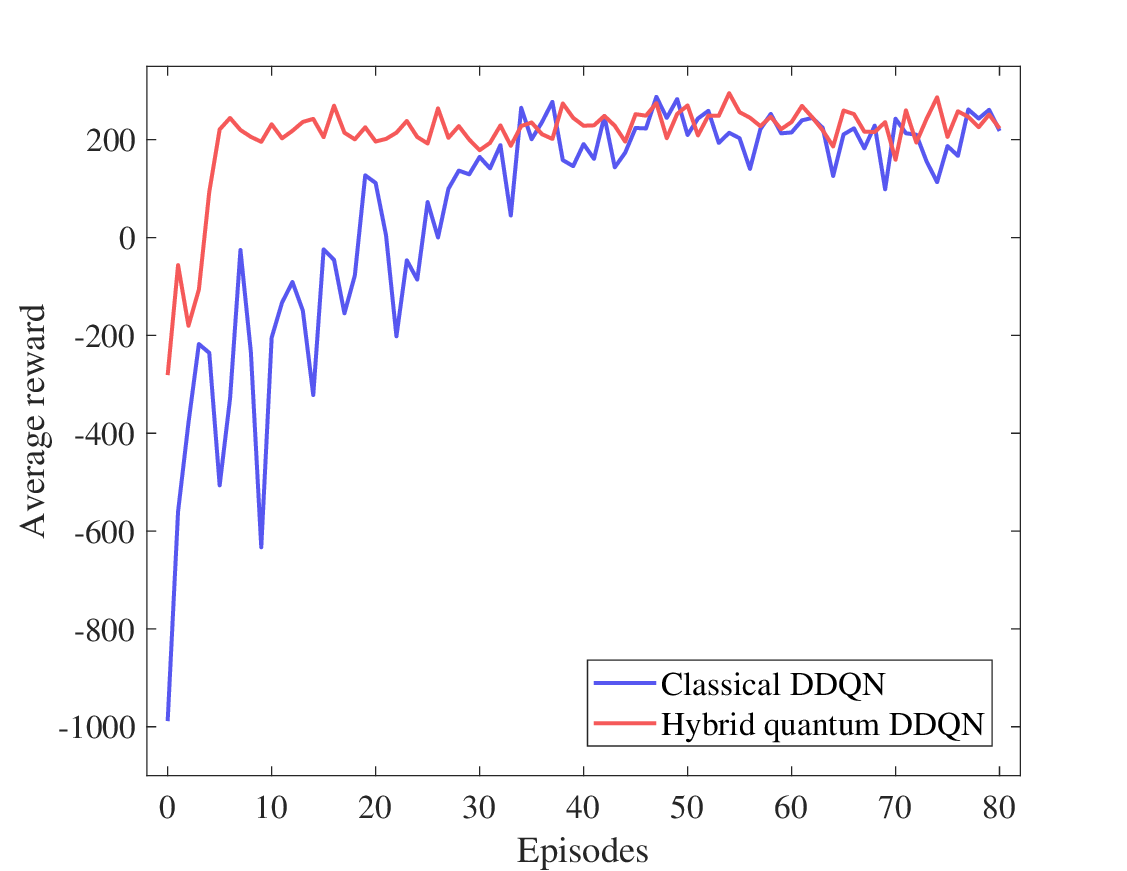}
      \label{figc1_APc}
    }
         \subfigure[{
         {$B_{\rm access}^{total}=60$}}MHz,$B_{\rm S}^{total}=110$MHz.]{
         \centering
         \includegraphics[width=1.6 in,height=1.5 in]{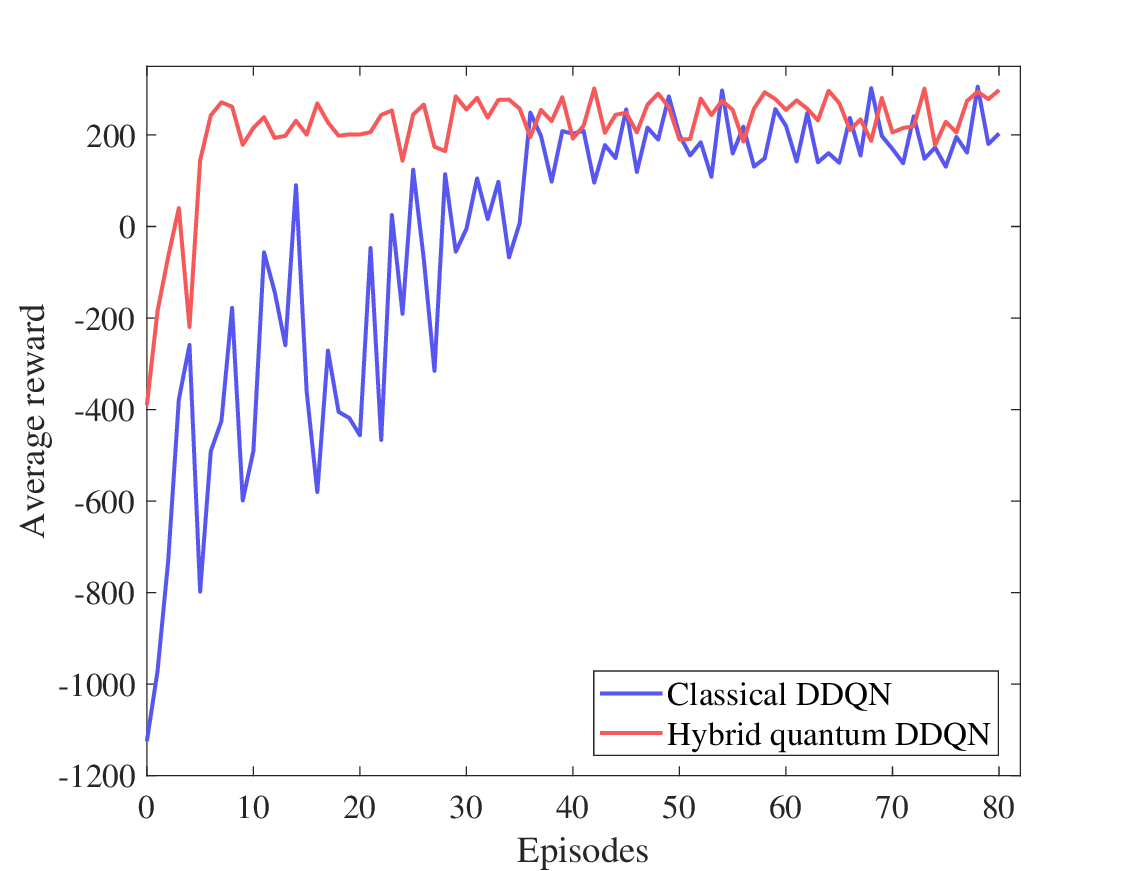}
      \label{figd1_APc}
    }
 \caption{The convergence of the proposed hybrid quantum DDQN in comparison with the classical counterpart.}
 \label{Fig_converg}
\end{figure}


Fig. 2 shows that the proposed hybrid quantum DDQN obtains a larger reward from a few data points and converges faster than the classical one, confirming that the hybrid quantum Q-network can accurately approximate the action-value function. The reason is that the hybrid architecture extracts both the harmonic and non-harmonic features from the data points~\cite{Kordzanganeh2023}.
\subsection{Efficiency}
 Two baseline algorithms are considered: i) Exhaustive approach for globally solving the discrete subproblem \eqref{eq10} with ADMM-based bandwidth allocation; and ii) The proposed hybrid quantum DDQN for solving \eqref{eq10} with equal bandwidth allocation.

\begin{figure}[htbp]
\centering
 \includegraphics[width=2.3 in,height=1.74in]{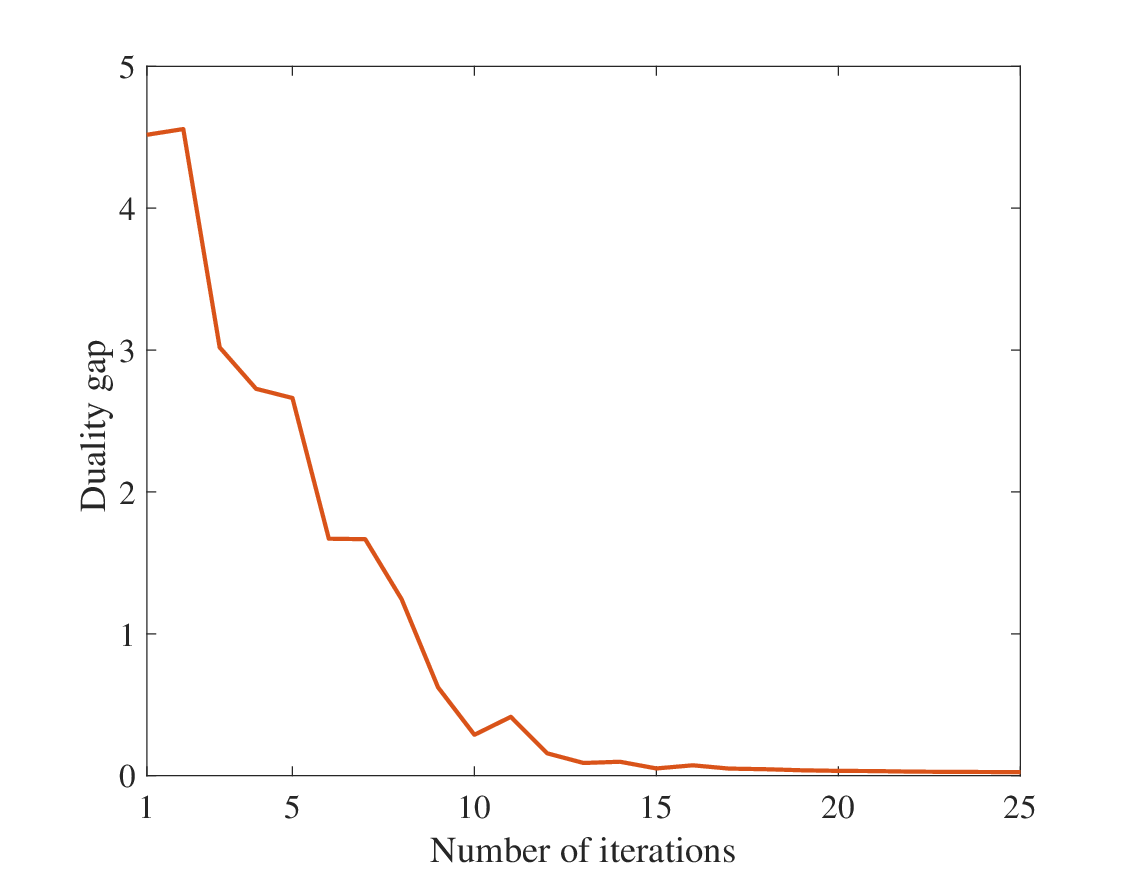}
\caption{Duality gap with {\footnotesize{$B_{\rm access}^{total}=50$}}MHz,$B_{\rm S}^{total}=100$MHz and $I=5\times10^5$bits.}
 \label{figa2_APc}
\end{figure}

Fig. \ref{figa2_APc} shows that the duality gap between the objective of primal  problem \eqref{eq6} and dual function \eqref{eq9}  is negligible under the proposed algorithm, which  means that the obtained dual optimum well approximates the solution of the primal problem \eqref{eq6}. The reason is that the proposed hybrid quantum DDQN can well solve the subproblem \eqref{eq10}. In fact, when the solution of subproblem \eqref{eq10} is globally optimal, our problem reduces to the convex problem w.r.t. bandwidth allocation.

\begin{figure}[htbp]
\centering
 \includegraphics[width=2.3 in,height=1.74in]{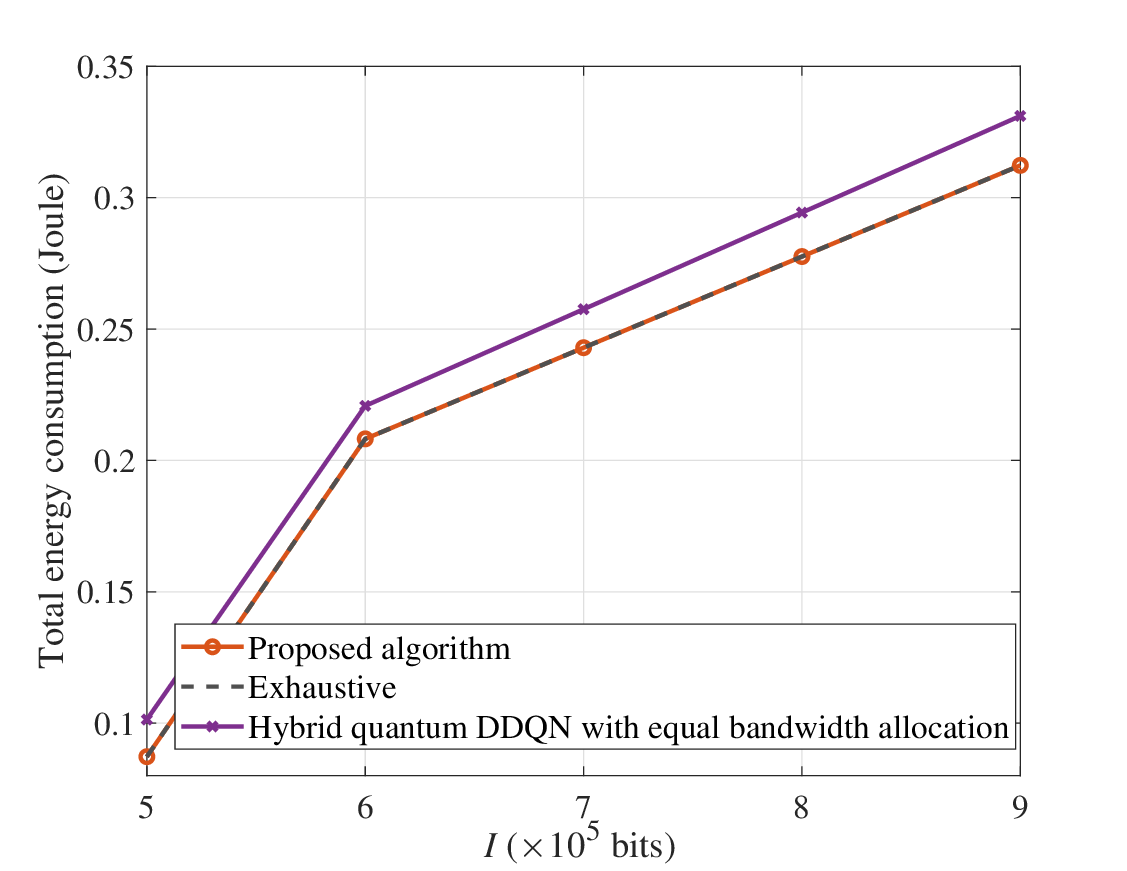}
\caption{Effect of the offloaded task size.}
 \label{figb2_APc}
\end{figure}

Fig. \ref{figb2_APc} shows that the use of the proposed algorithm can efficiently minimize the energy consumption for different sizes of the offloaded tasks, and also obtain the global optimum compared to the exhaustive approach. The joint design performs better  than the equal bandwidth allocation case.

\begin{figure}
     \centering
    \subfigure[Effect of the access bandwidth allocation with $B_{\rm S}^{total}=100$MHz.]{
         \centering
         \includegraphics[width=2.3 in,height=1.74 in]{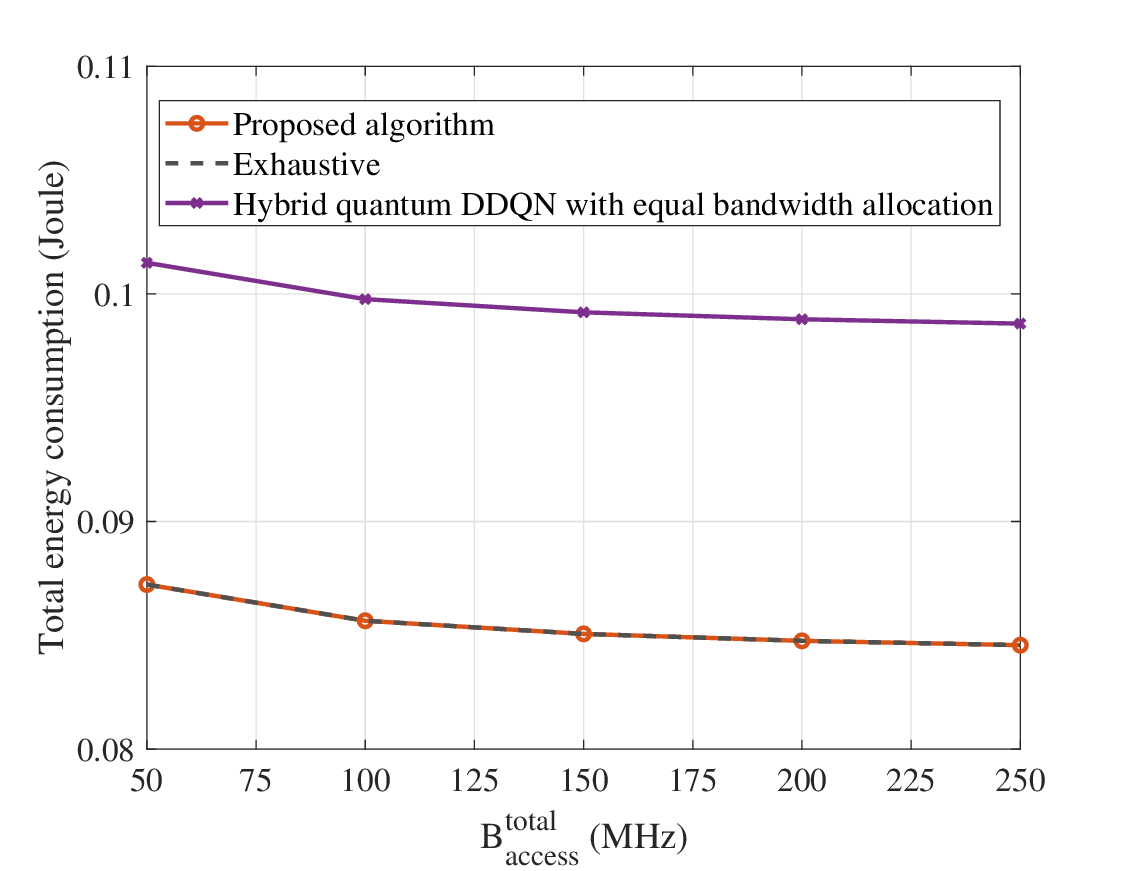}
      \label{figa3_APc}
     }
     \subfigure[Effect of the backhaul bandwidth allocation with {\footnotesize{$B_{\rm access}^{total}=50$}}MHz.]{
         \centering
         \includegraphics[width=2.3 in,height=1.74 in]{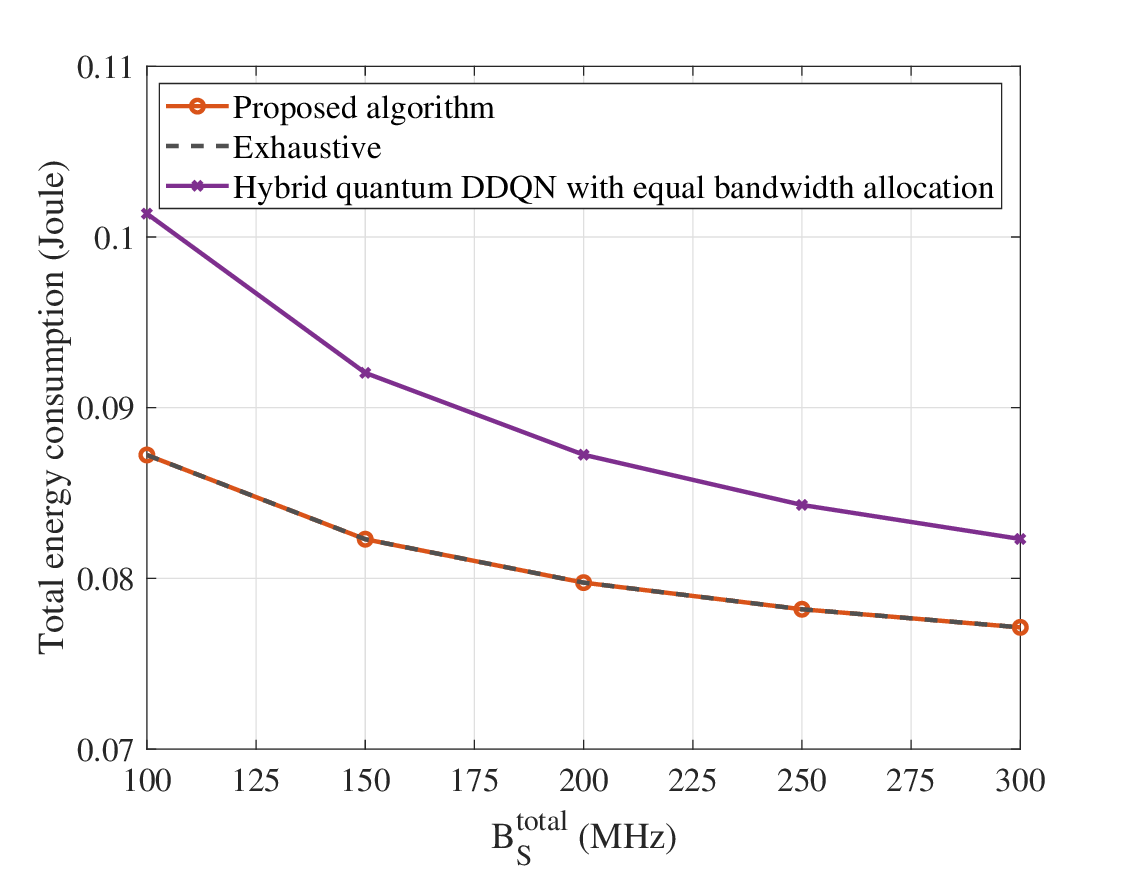}
      \label{figb3_APc}
    }
 \caption{The total energy consumption under different system  bandwidths for the access and backhaul links with $I=5\times10^5$bits.}
 \label{Fig_3}
\end{figure}

Figs. \ref{figa3_APc} and \ref{figb3_APc} show that the proposed algorithm can efficiently minimize the energy consumption under different total bandwidths for the access and backhaul links. Increasing the frequency bandwidths further reduces the total energy consumption since the communication delay is reduced and thus the energy consumption for task delivery is cut.

\section{Conclusions}\label{conclusion_section}
In the satellite-terrestrial networks, joint optimization of  the edge cloud selection and bandwidth allocation was studied, to minimize the total energy consumption of edge computing under delay and satellites' energy constraints. To efficiently solve it, an ADMM-inspired algorithm was proposed, and a novel hybrid quantum DDQN architecture was developed to solve the discrete subproblem. Numerical results confirmed that our algorithm could well approximate the global optimum and enable small duality gap. In addition, the proposed hybrid quantum DDQN could get large reward from a few data points.

\bibliographystyle{IEEEtran}

\begin{thebibliography}{}
\providecommand{\url}[1]{#1}
\csname url@samestyle\endcsname
\providecommand{\newblock}{\relax}
\providecommand{\bibinfo}[2]{#2}
\providecommand{\BIBentrySTDinterwordspacing}{\spaceskip=0pt\relax}
\providecommand{\BIBentryALTinterwordstretchfactor}{4}
\providecommand{\BIBentryALTinterwordspacing}{\spaceskip=\fontdimen2\font plus
\BIBentryALTinterwordstretchfactor\fontdimen3\font minus
  \fontdimen4\font\relax}
\providecommand{\BIBforeignlanguage}[2]{{%
\expandafter\ifx\csname l@#1\endcsname\relax
\typeout{** WARNING: IEEEtran.bst: No hyphenation pattern has been}%
\typeout{** loaded for the language `#1'. Using the pattern for}%
\typeout{** the default language instead.}%
\else
\language=\csname l@#1\endcsname
\fi
#2}}
\providecommand{\BIBdecl}{\relax}
\BIBdecl

\end{thebibliography}


\begin{thebibliography}{10}
\providecommand{\url}[1]{#1}
\csname url@samestyle\endcsname
\providecommand{\newblock}{\relax}
\providecommand{\bibinfo}[2]{#2}
\providecommand{\BIBentrySTDinterwordspacing}{\spaceskip=0pt\relax}
\providecommand{\BIBentryALTinterwordstretchfactor}{4}
\providecommand{\BIBentryALTinterwordspacing}{\spaceskip=\fontdimen2\font plus
\BIBentryALTinterwordstretchfactor\fontdimen3\font minus
  \fontdimen4\font\relax}
\providecommand{\BIBforeignlanguage}[2]{{%
\expandafter\ifx\csname l@#1\endcsname\relax
\typeout{** WARNING: IEEEtran.bst: No hyphenation pattern has been}%
\typeout{** loaded for the language `#1'. Using the pattern for}%
\typeout{** the default language instead.}%
\else
\language=\csname l@#1\endcsname
\fi
#2}}
\providecommand{\BIBdecl}{\relax}
\BIBdecl

\bibitem{Xiaoyan2020}
X.~Hu, L.~Wang, K.-K. Wong, M.~Tao, Y.~Zhang, and Z.~Zheng, ``Edge and central
  cloud computing: {A} perfect pairing for high energy efficiency and
  low-latency,'' \emph{{IEEE} Trans. Wireless Commun.}, vol.~19, no.~2, pp.
  1070--1083, Feb. 2020.

\bibitem{Alan-D}
A.~D. George and C.~M. Wilson, ``Onboard processing with hybrid and
  reconfigurable computing on small satellites,'' \emph{Proc. {IEEE}}, vol.
  106, no.~3, pp. 458--470, Mar. 2018.

\bibitem{Furong2023}
F.~Chai, Q.~Zhang, H.~Yao, X.~Xin, R.~Gao, and M.~Guizani, ``Joint multi-task
  offloading and resource allocation for mobile edge computing systems in
  satellite {IoT},'' \emph{{IEEE} Trans. Veh. Technol.}, vol.~72, no.~6, pp.
  7783--7795, June 2023.

\bibitem{Hassan_JSAC2024}
S.~S. Hassan, Y.~M. Park, Y.~K. Tun, W.~Saad, Z.~Han, and C.~S. Hong,
  ``Space{RIS}: {LEO} satellite coverage maximization in {6G Sub-THz} networks
  by {MAPPO DRL} and whale optimization,'' \emph{{IEEE} J. Sel. Areas Commun.},
  vol.~42, no.~5, pp. 1262--1278, May 2024.

\bibitem{Hangyu2024}
H.~Zhang, H.~Zhao, R.~Liu, A.~Kaushik, X.~Gao, and S.~Xu, ``Collaborative task
  offloading optimization for satellite mobile edge computing using multi-agent
  deep reinforcement learning,'' \emph{{IEEE} Trans. Veh. Technol.}, pp. 1--16,
  2024.

\bibitem{PennyLane}
\BIBentryALTinterwordspacing
V.~Bergholm \emph{et~al.}, ``{PennyLane}: {A}utomatic differentiation of hybrid
  quantum-classical computations,'' 2022. [Online]. Available:
  \url{https://arxiv.org/abs/1811.04968}
\BIBentrySTDinterwordspacing

\bibitem{Yen-Chi2020}
S.~Y.-C. Chen, C.-H.~H. Yang, J.~Qi, P.-Y. Chen, X.~Ma, and H.-S. Goan,
  ``Variational quantum circuits for deep reinforcement learning,'' \emph{IEEE
  Access}, vol.~8, pp. 141\,007--141\,024, 2020.

\bibitem{Narottama2022}
B.~Narottama and S.~Y. Shin, ``Quantum neural networks for resource allocation
  in wireless communications,'' \emph{{IEEE} Trans. Wireless Commun.}, vol.~21,
  no.~2, pp. 1103--1116, Feb. 2022.

\bibitem{Rainjonneau2023}
S.~Rainjonneau, I.~Tokarev, S.~Iudin, S.~Rayaprolu, K.~Pinto,
  D.~Lemtiuzhnikova, M.~Koblan, E.~Barashov, M.~Kordzanganeh, M.~Pflitsch, and
  A.~Melnikov, ``Quantum algorithms applied to satellite mission planning for
  earth observation,'' \emph{IEEE J. Sel. Top. Appl. Earth Obs. Remote Sens.},
  vol.~16, pp. 7062--7075, 2023.

\bibitem{Ansere2024}
J.~Adu~Ansere, D.~T. Tran, O.~A. Dobre, H.~Shin, G.~K. Karagiannidis, and T.~Q.
  Duong, ``Energy-efficient optimization for mobile edge computing with quantum
  machine learning,'' \emph{IEEE Wireless Commun. Lett.}, vol.~13, no.~3, pp.
  661--665, Mar. 2024.

\bibitem{huming}
M.~Hu, L.~Wang, B.~Tan, and S.~Jin, ``Two-tier 360-degree video delivery
  control in multiuser immersive communications systems,'' \emph{{IEEE} Trans.
  Veh. Technol.}, vol.~72, no.~3, pp. 4119--4123, Mar. 2023.

\bibitem{yang2016JSAC}
Y.~Yang, M.~Xu, D.~Wang, and Y.~Wang, ``Towards energy-efficient routing in
  satellite networks,'' \emph{{IEEE} J. Sel. Areas Commun.}, vol.~34, no.~12,
  pp. 3869--3886, Dec. 2016.

\bibitem{Feng-Wei}
Y.~Lin, W.~Feng, T.~Zhou, Y.~Wang, Y.~Chen, N.~Ge, and C.-X. Wang,
  ``Integrating satellites and mobile edge computing for {6G} wide-area edge
  intelligence: {M}inimal structures and systematic thinking,'' \emph{IEEE
  Netw.}, vol.~37, no.~2, pp. 14--21, Mar. 2023.

\bibitem{yuan2023}
Y.~Guo, E.~Faddoul, C.~Skouroumounis, and I.~Krikidis, ``{LEO} satellite-based
  space solar power systems,'' in \emph{IEEE ICASSP}, 2023, pp. 1--5.

\bibitem{zhiquanL2016}
M.~Hong, Z.-Q. Luo, and M.~Razaviyayn, ``Convergence analysis of alternating
  direction method of multipliers for a family of nonconvex problems,''
  \emph{SIAM J. Optimiz.}, vol.~26, no.~1, pp. 337--364, 2016.

\bibitem{laudecvpr2018}
E.~Laude, J.-H. Lange, J.~Schuepfer, C.~Domokos, L.~Leal-Taix\'{e}, F.~R.
  Schmidt, B.~Andres, and D.~Cremers, ``Discrete-continuous {ADMM} for
  transductive inference in higher-order {MRFs},'' in \emph{IEEE CVPR}, Feb.
  2018, pp. 1614--1624.

\bibitem{Santiago2023}
S.~Paternain, M.~Calvo-Fullana, L.~F.~O. Chamon, and A.~Ribeiro, ``Safe
  policies for reinforcement learning via primal-dual methods,'' \emph{IEEE
  Trans. Autom. Control}, vol.~68, no.~3, pp. 1321--1336, Mar. 2023.

\bibitem{huming2024}
M.~Hu, J.~Peng, L.~Wang, and K.-K. Wong, ``Scalable multiuser immersive
  communications with multi-numerology and mini-slot,'' \emph{{IEEE} Commun.
  Lett.}, vol.~28, no.~5, pp. 1201--1205, May 2024.

\bibitem{Hado2016}
H.~v. Hasselt, A.~Guez, and D.~Silver, ``Deep reinforcement learning with
  double {Q}-learning,'' in \emph{AAAI Conf.}, 2016, pp. 2094--2100.

\bibitem{Kordzanganeh2023}
M.~Kordzanganeh, D.~Kosichkina, and A.~Melnikov, ``Parallel hybrid networks:
  {A}n interplay between quantum and classical neural networks,'' \emph{Intell.
  Comput.}, vol.~2, p. 0028, 2023.

\end{thebibliography}

\end{document}